\documentclass[runningheads]{llncs}

\usepackage[T1]{fontenc}

\usepackage{booktabs}
\usepackage{graphicx,verbatim}

\begin{document}

\title{GANs vs. Diffusion Models for virtual staining with the HER2match dataset}

\author{Pascal Klöckner\inst{1} \and
José Teixeira\inst{1,2}\and
Diana Montezuma\inst{3} \and
Jaime S. Cardoso\inst{2} \and
Hugo M. Horlings\inst{1}\and
Sara P. Oliveira\inst{1}
}

\authorrunning{Klöckner et al.}

\institute{The Netherlands Cancer Institute, Amsterdam, The Netherlands \and
Faculty of Engineering, University of Porto, Porto, Portugal \and
IMP Diagnostics, Porto, Portugal}

\maketitle

\begin{abstract}
Virtual staining is a promising technique that uses deep generative models to recreate histological stains, providing a faster and more cost-effective alternative to traditional tissue chemical staining. Specifically for H\&E-HER2 staining transfer, despite a rising trend in publications, the lack of sufficient public datasets has hindered progress in the topic. Additionally, it is currently unclear which model frameworks perform best for this particular task. In this paper, we introduce the HER2match dataset, the first publicly available dataset with the same breast cancer tissue sections stained with both H\&E and HER2. Furthermore, we compare the performance of several Generative Adversarial Networks (GANs) and Diffusion Models (DMs), and implement a novel Brownian Bridge Diffusion Model for H\&E-HER2 translation. Our findings indicate that, overall, GANs perform better than DMs, with only the BBDM achieving comparable results. Furthermore, we emphasize the importance of data alignment, as all models trained on HER2match produced vastly improved visuals compared to the widely used consecutive-slide BCI dataset. This research provides a new high-quality dataset ([\textbf{available upon publication acceptance}]), improving both model training and evaluation. In addition, our comparison of frameworks offers valuable guidance for researchers working on the topic.


\end{abstract}

\section{Introduction}
\label{sec:intro} 
The field of pathology is undergoing a significant technological transformation~\cite{Huang_2025_IntelOnc_review}. The recent introduction and commercialization of generative Artificial Intelligence (AI) has also found its way into the field of digital pathology, with numerous new models and applications being published each year~\cite{Brodsky_2025_APLM_GenAIPath,Zhang_2025_arxiv_GenAIcpath}. One such technological advancement is virtual staining, a novel technique that enables image-to-image translation, \textit{i.e.} transforming a source domain to one or multiple target domains~\cite{Bai_2023_light,Latonen_2024_TB}. For example, we can convert autofluorescence (AF) images to Hematoxylin and Eosin (H\&E) images~\cite{Zhang_2022_IntelComp_AFtoHE} or H\&E images to immunohistochemistry (IHC) stains~\cite{Pati_2024_natureML_Multiplexer}. This technology promises benefits such as faster turnaround times, reduced costs, and the ability to stain multiple markers simultaneously, which might not be feasible with traditional methods for each sample. A particular well-suited use case of virtual staining is the determination of Human Epidermal Growth Factor Receptor 2 (HER2) status in breast cancer cases. This marker is crucial for treatment decisions and overall prognosis~\cite{Hacking_2022_Cancers_Biomarker} and is typically assessed by IHC. Since H\&E staining is cost-effective, fast, and routinely performed on all breast cancer samples, generating virtual HER2 samples from it is a promising avenue.

\subsubsection{Related work in HER2 virtual staining.}
In 2022, Liu \textit{et al.}~\cite{Liu_2022_CVPR_BCI} published the first work attempting HER2 virtual staining from H\&E images, together with the first public set of data for this task: the Breast Cancer Immunohistochemistry (BCI) dataset. While most subsequent models are based on this dataset, it contains several artifacts, and due to the staining of H\&E and IHC on consecutive tissue sections, image pairs do not align well. Although most regions may correspond, the same cells are not present in both images, which makes model training more difficult. Some models try to address these imperfect pairings by employing specialized architectures, but, in the end, high-quality datasets are essential to advance virtual staining of HER2.

Another factor affecting the quality of virtually stained images is the performance of the model itself. Several frameworks, such as Generative Adversarial Networks (GANs) or Diffusion Models (DMs), are commonly used for these tasks, with GANs being the most popular for HER2 virtual staining from H\&E, but DMs gaining momentum~\cite{Klockner_2025_npjDigMed}. However, it remains unclear which framework is better for this specific task, and thus, we evaluate three commonly used GANs and three DMs in this study (Figure~\ref{fig:model_architectures}).

\subsubsection{Generative Adversarial Networks.}
As the name suggests, GANs employ an adversarial process to guide model training~\cite{Goodfellow_2014_arxiv_GAN}. The generator (G) takes in the source image (\textit{e.g.}, an H\&E image) and generates a virtual image with the style of the target domain (\textit{e.g.}, IHC). A second model, the discriminator (D), is responsible for distinguishing between fake and real images from the target domain, encouraging the generator to produce more realistic images.

\begin{figure}[t]
    \centering
    \includegraphics[width=\textwidth]{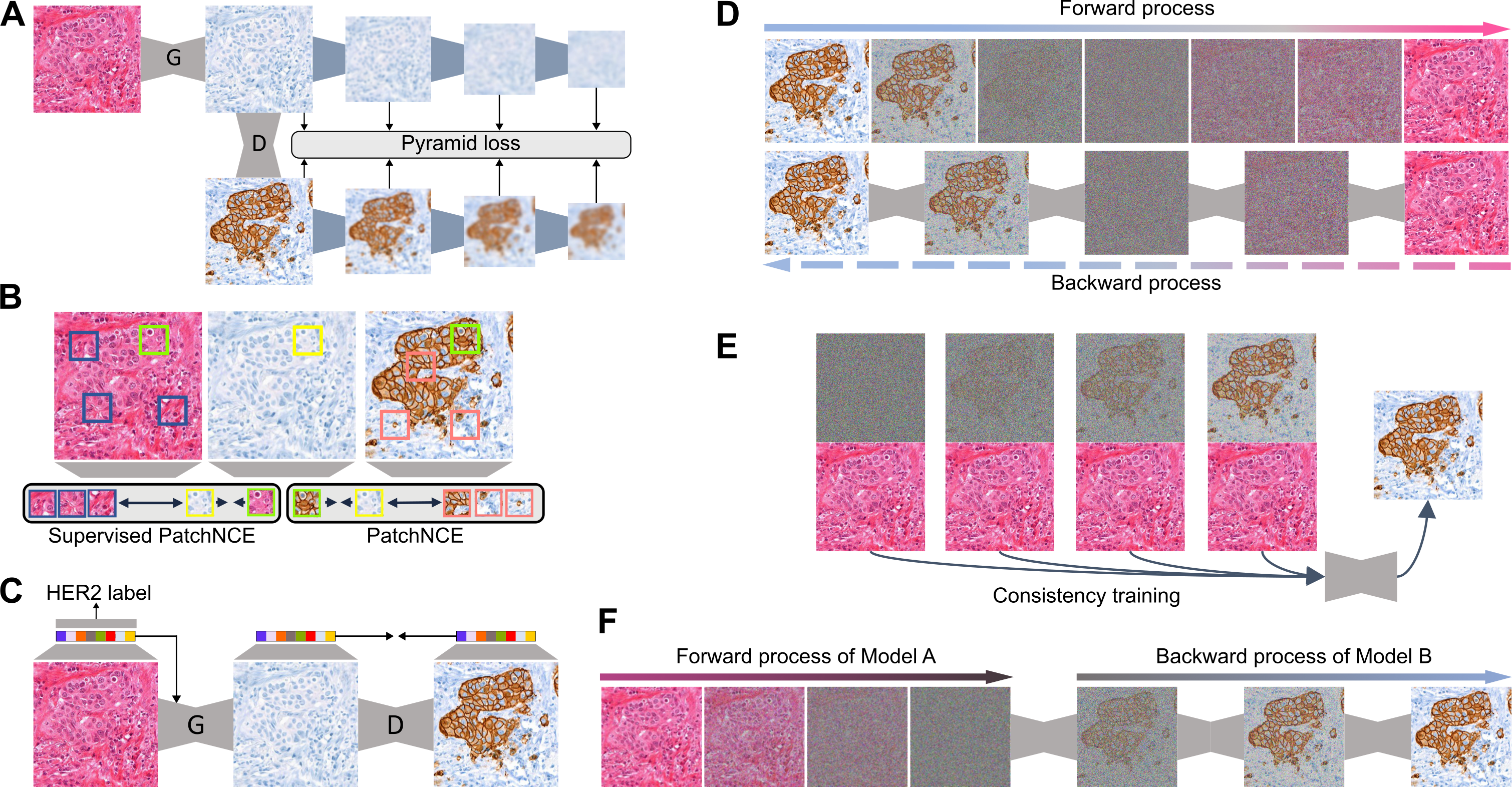}
    \caption{H\&E-to-IHC virtual staining frameworks: (A) Pyramid pix2pix, (B) ASP, (C) BCIstainer, (D) DDIB, (E) CM and (F) BBDM.}
    \label{fig:model_architectures}
\end{figure}

Pyramid pix2pix (ppx2px)~\cite{Liu_2022_CVPR_BCI} is one of the standard GAN-based models used for virtual staining tasks and serves as the base building block to many other models. Inspired by the widely successful Pix2Pix~\cite{Isola_2017_IEEE_Pix2Pix} model, it accounts for slight image registration imperfections by computing the L1 loss on multiple scales of the smoothed output image (pyramidal loss, as in Figure~\ref{fig:model_architectures}A). This approach ensures that images that are not perfectly pixel (px) aligned and might not even contain the same cells (as is for consecutive sections), can still be used for training. The model is trained by optimizing a combination of the standard conditional GAN loss, L1 loss, and the pyramidal loss (four different scales with equal weighting).

Based on the pyramid Pix2Pix architecture, Li~\textit{et al.}~\cite{Li_2023_MICCAI_ASP} introduced a novel Adaptive Supervised PatchNCE (ASP) loss function to also address the source-target pixel mismatch (Figure ~\ref{fig:model_architectures}B). This contrastive learning loss function is based on the PatchNCE loss~\cite{Park_ECCV_2020_CUT} but applied to both input-generated and target-generated image pairs. Additionally, the loss is weighted by the cosine similarity of the embeddings of the anchor tiles in both the generated and ground-truth images, which accounts for regions where there is less correspondence.

Another model based on the GAN framework is the BCIstainer model~\cite{Zhu_2023_arxiv_BCIchallenge_BCIstainer}. Despite being less widely used compared to the previously mentioned models, it achieved second place in the BCI Grand Challenge~\cite{Zhu_2023_arxiv_BCIchallenge_BCIstainer} and demonstrated the best qualitative results. In addition to the standard GAN loss, it calculates the mean absolute error (MAE) and includes a loss based on the Structural Similarity Index Measure (SSIM). Furthermore, it has a classification module that uses embeddings to condition the backbone of the generator (Figure ~\ref{fig:model_architectures}C).
    
\subsubsection{Diffusion Models.}
Diffusion models are generative models that begin with noise and, through an iterative denoising process, end up with a generated image. The forward pass, or diffusion process, involves adding a small amount of Gaussian noise to an image, and then the model learns the backward pass, or denoising process. These steps ultimately allow us to sample from Gaussian noise to generate new samples with a desired target distribution. There are several classes of diffusion models, such as Denoising Diffusion Probabilistic Models (DDPM)~\cite{Ho_2020_arxiv_DDPM} or Denoising Diffusion Implicit Models (DDIM)~\cite{Song_2022_arxiv_DDIM}, which mostly differ in their formulation of the diffusion and denoising processes. 

Dual Diffusion Implicit Bridges (DDIB)~\cite{Su_2023_ICLR_DDIB} are the basis of virtual staining methods such as PST-DIFF~\cite{He_2024_IEEETMI_PSTDIFF}. They can be thought of as a concatenation of two DDIMs, with each model independently trained on its respective domain (Figure ~\ref{fig:model_architectures}D). During inference, the source-to-latent encoding can be used as an input for the latent-to-target decoder, enabling deterministic pairwise translation from the source to the target domain. The ability to train each domain individually is particularly desirable, as it allows for the translation between multiple different domains without training a single model for each possible pairing.

Consistency Models (CM)~\cite{Song_2023_arxiv_consistencymodels} are a family of models designed to alleviate the slow sampling procedure of DMs (often requiring up to 1000 iterations per sample generation) by providing single-step (or a significantly reduced step) inference. They can be distilled from previously trained DMs or trained as standalone models, ensuring consistent outcomes regardless of the sample point in the data-to-noise trajectory, which ultimately enables single-step inference (Figure ~\ref{fig:model_architectures}E). While efficient inference time is not as important for virtual staining as in other fields, it is still desirable, especially when considering multiple virtual stains for gigapixel images such as WSIs. In the context of virtual staining, conditional CMs have been developed by concatenating the source image to the sampled noise, enabling pairwise source-to-target translation~\cite{Bhagat_2025_arxiv_conditionalconsistencyguidedimage}.

While accounting for paired data in most DMs architectures is done via conditioning, Brownian Bridge Diffusion Models (BBDMs) inherently integrate both the source and the target domain images~\cite{Li_2023_CVPR_BBDM}. Instead of going from the source to Gaussian noise, BBDMs are trained by going from the source to the target directly, with additional noise injection (Figure ~\ref{fig:model_architectures}F). The backbone model is therefore not just predicting added noise at each step, but also the gradient from the source to the target at each step. BBDM have demonstrated success in other virtual staining tasks, such as transforming lower-resolution autofluorescence images into high-resolution H\&E images~\cite{Zhang_2025_NatureComms_BBDM}. To our knowledge, we are the first to implement this model class for HER2 virtual staining from H\&E.

\subsubsection{Scope.}
The currently available datasets, such as BCI, are very useful for advancing virtual staining techniques, but fall short for a detailed model training and evaluation, since they consist of images from consecutive tissue samples, lacking an accurate cell-to-cell matching in the H\&E and IHC pairs. In this paper, we introduce the HER2match dataset \textbf{[available upon publication acceptance]}, a high-resolution, manually curated set of H\&E and IHC (HER2) paired tiles from the same tissue slide. Additionally, we manually cleaned the BCI dataset, resulting in the BCI-clean set, which contains only tiles without (major) artifacts. From the model point of view, we compare several GAN- and DM-based architectures and evaluate their performance, including the first implementation of a BBDM for H\&E to HER2 virtual staining. In summary, the main contributions of this study include:

\begin{enumerate}
\item The first publicly available dataset containing paired images of same-slide H\&E and HER2 stainings from breast cancer cases;
\item The first implementation of a Brownian Bridge Diffusion Model in the context of virtual staining of HER2 from H\&E;
\item   A performance evaluation comparing DMs and GANs for virtual staining tasks across datasets of varying quality.
\end{enumerate}

\section{Methods}
\label{sec:methods} 
We train and test three GAN architectures (Pyramid px2px, ASP, and BCIstainer) used for paired datasets as well as three DM architectures (DDIB, CM, and BBDM) on the BCI dataset, a cleaned version of the BCI dataset (BCI-clean), as well as our own HER2match dataset.

\subsubsection{BCI dataset.}
\begin{figure}[b]
    \centering
    \includegraphics[width=\textwidth]{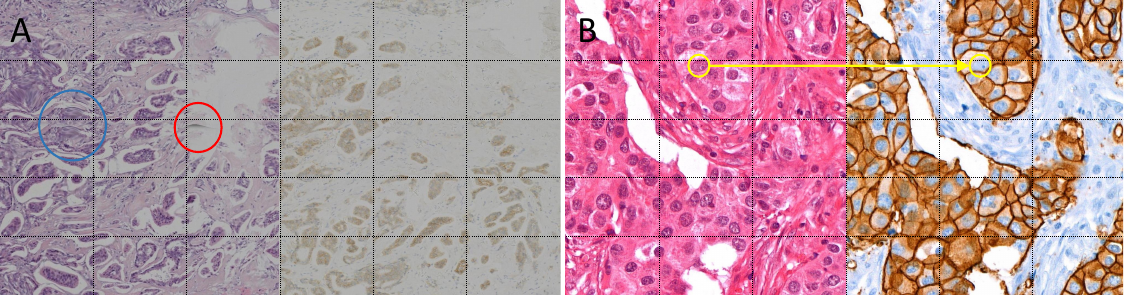}
    \caption{Examples from the BCI (A) and HER2match (B) datasets. Blue and red circles highlight two types of registration artifacts commonly present in BCI images, and yellow circles depict the level of alignment in the HER2match pairs.}
    \label{fig:dataset_comparison}
\end{figure}

The BCI dataset~\cite{Liu_2022_CVPR_BCI} consists of 4873 tile pairs (1024x1024 px, with a resolution of 0.46µm/px) from 51 consecutive H\&E samples and their corresponding HER2 slides. The dataset is split into 3896 images for training and 977 images for validation, stratified by case. The dataset contains some artifacts, such as doubling of the image sections or dark/black shades (Figure~\ref{fig:dataset_comparison}A). Thus, we manually assessed all tiles and cleaned the dataset, resulting in the BCI-clean version with 1112 images in the training set and 144 images in the validation set (the lists are available at \textbf{[available upon publication acceptance]}).

\subsubsection{HER2match dataset.}
The HER2match dataset consists of tiles from WSIs (n=17 cases) that were stained with H\&E, washed out, and re-stained with HER2 IHC (Figure~\ref{fig:dataset_comparison}B). All slides were scanned with a 3DHistech P1000 scanner, at 40X magnification, with a resolution of 0.25 µm/px. To ensure that the scoring was not affected by this protocol (as this might lower the amount of chromogen deposition in tissue), the final IHC slides were reviewed by a pathologist to confirm the original diagnostic HER2 score (0, 1+, 2+, and 3+). For automatic registration of the H\&E and IHC slides, we used the DeeperHistReg framework~\cite{Wodzinski_2024_arxiv_DeeperHistReg}, with their non-rigid high-resolution pipeline to transform the IHC slides (in contrast to BCI which transforms the H\&E images). Two WSIs required semi-automatic registration using Warpy~\cite{Chiaruttini_2022_FrontiersCS_Warpy}, as the initial registration did not produce satisfying results. After successful registration, tissue areas were determined using Otsu thresholding~\cite{Bangare_2015_IJAER_otsu}, and regions containing mostly tumor were manually selected. The resulting ROIs were then cropped into 1024x1024 px tiles at the highest resolution, resulting in 36,209 H\&E-HER2 tile pairs. These tiles were manually curated to account for tile-level artifacts, resulting in a final set of 21,172 tile pairs. The images were stratified by case, ensuring that the training (11,610 pairs), validation (3,582 pairs), and test (5,980 pairs) sets included at least one case from each HER2 score category.

\subsubsection{Model implementation.}
Pyramid pix2pix, ASP, BCIstainer, DDIB, and CM were all trained with the hyperparameters specified in the corresponding GitHub repositories~\cite{pp2p_git,asp_git,bcistainer_git,ddib_git,cm_git}. The BBDM was trained for 500 epochs with a U-net backbone. Since the BCI and HER2match datasets have different resolutions, the BCI tiles were cropped into four smaller tiles and the HER2match tiles were rescaled to half of their original size to achieve similar resolution ($\approx0.5$ µm/px) and tile size (512x512 px) for both datasets, ensuring that those parameters do not affect the model comparisons. All models were trained on NVIDIA A100/H100 GPUs. For evaluation, we calculate classic image metrics such as SSIM~\cite{Hore_2010_CPR_PSNR} and Peak Signal-to-Noise Ratio (PSNR)~\cite{Hore_2010_CPR_PSNR}, as well as metrics based on deep neural networks such as Fréchet inception distance (FID)~\cite{Heusel_2017_arxiv_FID}, Kernel inception distance (KID)~\cite{Bińkowski_2021_arxiv_KID)}, and the Learned Perceptual Image Patch Similarity (LPIPS)~\cite{Zhang_2018_CVPR_LPIPS}. 

\subsubsection{Linear mixed effect modeling.}
We are interested in evaluating the performance of both the GAN and DMs frameworks on the most widely used virtual staining dataset, before and after excluding tiles with artifacts. To unravel the interaction effect of the framework (GAN vs. DM) and dataset (BCI vs. BCI-clean), we construct linear mixed effects models using the lme4 package in R~\cite{Bates_2015_JSS_lme4}  for each of the pair-wise image metrics (SSIM, PSNR, and LPIPS). In our models, we treat the framework and dataset as fixed effects, while accounting for the specific models and individual images as random effects, resulting in equation \ref{eqn:linear_model}:
\begin{equation}
\label{eqn:linear_model}
Metric = Framework \cdot Dataset + (1|Model) + (1|Image)
\end{equation}
\section{Results}
\label{sec:results} 
\begin{figure}[t]
    \centering
    \includegraphics[width=1\linewidth]{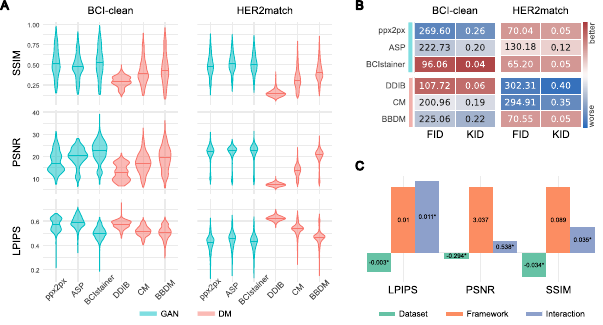}
    \caption{(A) Distribution of SSIM, PSNR, and LPIPS on the test sets. (B) Heatmap of FID and KID. (C) Results of the Linear models, with * denoting p$\leq$0.001.}
    \label{fig:violinplots}
\end{figure}

While the GAN model demonstrates the best performance across nearly all metrics (Figure~\ref{fig:violinplots}A \& B), the results of the linear mixed effects modeling (Figure~\ref{fig:violinplots}C), reveal that only the dataset and interaction effects are significant. Training on BCI-clean vs. BCI has a significant negative effects on SSIM, MS-SSIM and PSNR while the interaction effect for GANs mostly counteracts this impact and even provides an advantage in PSNR. In contrast, training on BCI-clean yields a positive effect for LPIPS, and DMs benefit from training on the cleaned dataset. Overall, BCIstainer ranks as the best or second-best model.

\begin{figure}[t]
    \centering
    \includegraphics[width=\textwidth]{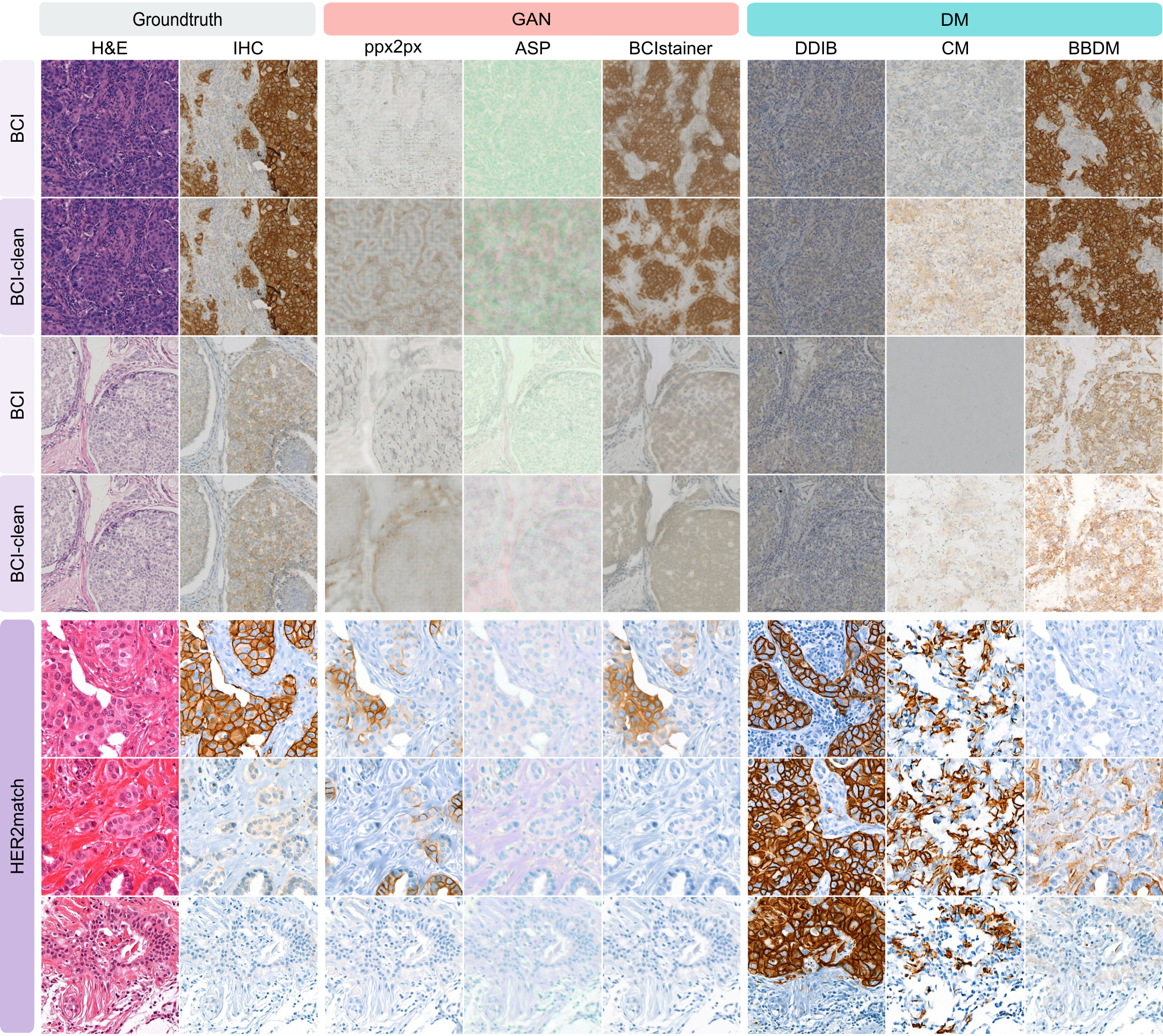}
    \caption{Image examples for models trained on BCI, BCI-clean, and HER2match.}
    \label{fig:Images_BCI_BCI-clean}
\end{figure}

An examination of the generated images (Figure~\ref{fig:Images_BCI_BCI-clean}) shows that GANs generally capture the tissue structures more effectively, while DMs look crisper and occasionally more realistic, even though DDIB and CM do not generate images with the corresponding tissue structures found in H\&E staining. Based on the previously mentioned metrics, the model that best captures both morphology and staining is BCIstainer, which, nevertheless, does not yet have sufficient quality for clinical practice. Focusing on the HER2match dataset, GANs and BBDM excel at mimicking morphology, while both DDIB and CM still take some parts of the source image into account, although not consistently, and often produce a uniform style for any input, such as strong membrane staining. Pyramid pix2pix, BCIstainer, and BBDM are able to generate images with the target style, but have difficulty with representing the staining pattern accurately.

\section{Discussion}
\label{sec:discussion} 
In this paper, we compared different DM- and GAN-based models on the BCI dataset, a cleaned version of the BCI dataset, as well as our own HER2match dataset. Although DMs have superseded GANs in other fields, our results indicate that this may not yet be the case for virtual staining. On average, GANs outperformed DMs across all metrics for all datasets and also produced better visual results, particularly in preserving morphology details. One important factor that might make GANs better suited for virtual staining tasks is their capability to generate actual images at each step. This allows for the integration of training objectives that take pathological information (such as the chromogen distribution in the image) into account. While GANs are more challenging to train due to the need to balance the generator and discriminator dynamics, this ability to include specialized loss functions might give them an edge over DM in the context of virtual staining. 

Our study also emphasized the importance of proper data alignment and that it is not just enough to have paired consecutive slides without artifacts. We believe that the introduction of our new HER2match dataset will significantly advance research in the topic, as it is the only publicly available dataset with cell-to-cell matching image pairs from H\&E and HER2 IHC-stained WSIs. This enables a more precise evaluation of the validity of virtual staining models, as we can now assess if specific cells are accurately stained. In this context, it also highlights the lack of specialized metrics, suggesting that future research should invest in creating specific perceptual metrics for virtual staining, for example, by adapting LPIPS for histology images.

Finally, we presented the first application of a Brownian Bridge Diffusion Model for H\&E-HER2 staining translation. The BBDM was the only DM that demonstrated comparable visual quality to the GANs, showcasing its ability to preserve morphology and transfer the staining style. However, while both GANs and BBDM produced realistic images resembling IHC, neither was consistently achieving appropriate staining levels. Future work must focus on improving pathological fidelity, as current models are not yet ready for clinical use.

 \begin{credits}
 \subsubsection{\ackname} The authors would like to thank Bart de Rooij, Simone van Veldhoven, and Joyce Sanders. This publication is part of the project ``stAIns: AI-based virtual immuno staining from H\&E slides'' with file number NGF.1609.241.018 of the research programme NGF AiNED XS Europa 2024, which is (partly) financed by the Dutch Research Council (NWO).

\subsubsection{\discintname}
All authors declare no financial or non-financial competing interests.
\end{credits}

\bibliographystyle{naturemag} 
\bibliography{references.bib}

\begin{thebibliography}{10}
\expandafter\ifx\csname url\endcsname\relax
  \def\url#1{\texttt{#1}}\fi
\expandafter\ifx\csname urlprefix\endcsname\relax\def\urlprefix{URL }\fi
\providecommand{\bibinfo}[2]{#2}
\providecommand{\eprint}[2][]{\url{#2}}

\bibitem{Huang_2025_IntelOnc_review}
\bibinfo{author}{Huang, Q.}, \bibinfo{author}{Wu, S.}, \bibinfo{author}{Ou, Z.} \& \bibinfo{author}{Gao, Y.}
\newblock \bibinfo{title}{Computational pathology: A comprehensive review of recent developments in digital and intelligent pathology}.
\newblock \emph{\bibinfo{journal}{Intelligent Oncology}} \textbf{\bibinfo{volume}{1}}, \bibinfo{pages}{139--159} (\bibinfo{year}{2025}).
\newblock \urlprefix\url{https://www.sciencedirect.com/science/article/pii/S2950261625000214}.

\bibitem{Brodsky_2025_APLM_GenAIPath}
\bibinfo{author}{Brodsky, V.} \emph{et~al.}
\newblock \bibinfo{title}{Generative artificial intelligence in anatomic pathology}.
\newblock \emph{\bibinfo{journal}{Archives of Pathology \& Laboratory Medicine}} \textbf{\bibinfo{volume}{149}}, \bibinfo{pages}{298--318} (\bibinfo{year}{2025}).
\newblock \urlprefix\url{https://doi.org/10.5858/arpa.2024-0215-RA}.

\bibitem{Zhang_2025_arxiv_GenAIcpath}
\bibinfo{author}{Zhang, Y.} \emph{et~al.}
\newblock \bibinfo{title}{Generative models in computational pathology: A comprehensive survey on methods, applications, and challenges} (\bibinfo{year}{2025}).
\newblock \urlprefix\url{https://arxiv.org/abs/2505.10993}.
\newblock \eprint{2505.10993}.

\bibitem{Bai_2023_light}
\bibinfo{author}{Bai, B.} \emph{et~al.}
\newblock \bibinfo{title}{Deep learning-enabled virtual histological staining of biological samples}.
\newblock \emph{\bibinfo{journal}{Light: Science \& Applications}} \textbf{\bibinfo{volume}{12}}, \bibinfo{pages}{57} (\bibinfo{year}{2023}).

\bibitem{Latonen_2024_TB}
\bibinfo{author}{Latonen, L.}, \bibinfo{author}{Koivukoski, S.}, \bibinfo{author}{Khan, U.} \& \bibinfo{author}{Ruusuvuori, P.}
\newblock \bibinfo{title}{Virtual staining for histology by deep learning}.
\newblock \emph{\bibinfo{journal}{Trends in Biotechnology}} \textbf{\bibinfo{volume}{42}}, \bibinfo{pages}{1177–1191} (\bibinfo{year}{2024}).

\bibitem{Zhang_2022_IntelComp_AFtoHE}
\bibinfo{author}{Zhang, Y.} \emph{et~al.}
\newblock \bibinfo{title}{Virtual staining of defocused autofluorescence images of unlabeled tissue using deep neural networks}.
\newblock \emph{\bibinfo{journal}{Intelligent Computing}} \textbf{\bibinfo{volume}{2022}} (\bibinfo{year}{2022}).
\newblock \urlprefix\url{https://spj.science.org/doi/abs/10.34133/2022/9818965}.

\bibitem{Pati_2024_natureML_Multiplexer}
\bibinfo{author}{Pati, P.} \emph{et~al.}
\newblock \bibinfo{title}{Accelerating histopathology workflows with generative ai-based virtually multiplexed tumour profiling}.
\newblock \emph{\bibinfo{journal}{Nature Machine Intelligence}} \textbf{\bibinfo{volume}{6}}, \bibinfo{pages}{1077–1093} (\bibinfo{year}{2024}).
\newblock \urlprefix\url{http://dx.doi.org/10.1038/s42256-024-00889-5}.

\bibitem{Hacking_2022_Cancers_Biomarker}
\bibinfo{author}{Hacking, S.~M.}, \bibinfo{author}{Yakirevich, E.} \& \bibinfo{author}{Wang, Y.}
\newblock \bibinfo{title}{From immunohistochemistry to new digital ecosystems: A state-of-the-art biomarker review for precision breast cancer medicine}.
\newblock \emph{\bibinfo{journal}{Cancers}} \textbf{\bibinfo{volume}{14}} (\bibinfo{year}{2022}).
\newblock \urlprefix\url{https://www.mdpi.com/2072-6694/14/14/3469}.

\bibitem{Liu_2022_CVPR_BCI}
\bibinfo{author}{Liu, S.} \emph{et~al.}
\newblock \bibinfo{title}{{BCI: Breast Cancer Immunohistochemical Image Generation through Pyramid Pix2pix}}.
\newblock In \emph{\bibinfo{booktitle}{{2022 IEEE/CVF Conference on Computer Vision and Pattern Recognition Workshops -- CVPRW}}}, \bibinfo{pages}{1814--1823} (\bibinfo{year}{2022}).

\bibitem{Klockner_2025_npjDigMed}
\bibinfo{author}{Klöckner, P.} \emph{et~al.}
\newblock \bibinfo{title}{{H\&E to IHC virtual staining methods in breast cancer: an overview and benchmarking}}.
\newblock \emph{\bibinfo{journal}{NPJ Digital Medicine}}  (\bibinfo{year}{2025}).

\bibitem{Goodfellow_2014_arxiv_GAN}
\bibinfo{author}{Goodfellow, I.~J.} \emph{et~al.}
\newblock \bibinfo{title}{Generative adversarial networks} (\bibinfo{year}{2014}).
\newblock \urlprefix\url{https://arxiv.org/abs/1406.2661}.
\newblock \eprint{1406.2661}.

\bibitem{Isola_2017_IEEE_Pix2Pix}
\bibinfo{author}{Isola, P.}, \bibinfo{author}{Zhu, J.-Y.}, \bibinfo{author}{Zhou, T.} \& \bibinfo{author}{Efros, A.~A.}
\newblock \bibinfo{title}{{Image-to-Image Translation with Conditional Adversarial Networks}}.
\newblock In \emph{\bibinfo{booktitle}{{2017 IEEE Conference on Computer Vision and Pattern Recognition -- CVPR}}}, \bibinfo{pages}{5967--5976} (\bibinfo{year}{2017}).

\bibitem{Li_2023_MICCAI_ASP}
\bibinfo{author}{Li, F.}, \bibinfo{author}{Hu, Z.}, \bibinfo{author}{Chen, W.} \& \bibinfo{author}{Kak, A.}
\newblock \bibinfo{title}{{{Adaptive Supervised PatchNCE Loss for Learning H{\&}E-to-IHC Stain Translation with Inconsistent Groundtruth Image Pairs}}}.
\newblock In \emph{\bibinfo{booktitle}{Medical Image Computing and Computer Assisted Intervention -- MICCAI}}, \bibinfo{pages}{632--641} (\bibinfo{year}{2023}).

\bibitem{Park_ECCV_2020_CUT}
\bibinfo{author}{Park, T.}, \bibinfo{author}{Efros, A.~A.}, \bibinfo{author}{Zhang, R.} \& \bibinfo{author}{Zhu, J.-Y.}
\newblock \bibinfo{title}{Contrastive learning for unpaired image-to-image translation}.
\newblock In \bibinfo{editor}{Vedaldi, A.}, \bibinfo{editor}{Bischof, H.}, \bibinfo{editor}{Brox, T.} \& \bibinfo{editor}{Frahm, J.-M.} (eds.) \emph{\bibinfo{booktitle}{European Conference on Computer Vision -- ECCV 2020}}, \bibinfo{pages}{319--345} (\bibinfo{publisher}{Springer International Publishing}, \bibinfo{year}{2020}).

\bibitem{Zhu_2023_arxiv_BCIchallenge_BCIstainer}
\bibinfo{author}{Zhu, C.} \emph{et~al.}
\newblock \bibinfo{title}{Breast cancer immunohistochemical image generation: a benchmark dataset and challenge review} (\bibinfo{year}{2023}).
\newblock \urlprefix\url{https://arxiv.org/abs/2305.03546}.
\newblock \eprint{2305.03546}.

\bibitem{Ho_2020_arxiv_DDPM}
\bibinfo{author}{Ho, J.}, \bibinfo{author}{Jain, A.} \& \bibinfo{author}{Abbeel, P.}
\newblock \bibinfo{title}{Denoising diffusion probabilistic models} (\bibinfo{year}{2020}).
\newblock \urlprefix\url{https://arxiv.org/abs/2006.11239}.
\newblock \eprint{2006.11239}.

\bibitem{Song_2022_arxiv_DDIM}
\bibinfo{author}{Song, J.}, \bibinfo{author}{Meng, C.} \& \bibinfo{author}{Ermon, S.}
\newblock \bibinfo{title}{Denoising diffusion implicit models} (\bibinfo{year}{2022}).
\newblock \urlprefix\url{https://arxiv.org/abs/2010.02502}.
\newblock \eprint{2010.02502}.

\bibitem{Su_2023_ICLR_DDIB}
\bibinfo{author}{Su, X.}, \bibinfo{author}{Song, J.}, \bibinfo{author}{Meng, C.} \& \bibinfo{author}{Ermon, S.}
\newblock \bibinfo{title}{Dual diffusion implicit bridges for image-to-image translation}.
\newblock In \emph{\bibinfo{booktitle}{International Conference on Learning Representations}} (\bibinfo{year}{2023}).

\bibitem{He_2024_IEEETMI_PSTDIFF}
\bibinfo{author}{He, Y.} \emph{et~al.}
\newblock \bibinfo{title}{{PST-Diff: Achieving High-Consistency Stain Transfer by Diffusion Models With Pathological and Structural Constraints}}.
\newblock \emph{\bibinfo{journal}{IEEE Trans. Med. Imaging}} \textbf{\bibinfo{volume}{43}}, \bibinfo{pages}{3634--3647} (\bibinfo{year}{2024}).

\bibitem{Song_2023_arxiv_consistencymodels}
\bibinfo{author}{Song, Y.}, \bibinfo{author}{Dhariwal, P.}, \bibinfo{author}{Chen, M.} \& \bibinfo{author}{Sutskever, I.}
\newblock \bibinfo{title}{Consistency models} (\bibinfo{year}{2023}).
\newblock \urlprefix\url{https://arxiv.org/abs/2303.01469}.
\newblock \eprint{2303.01469}.

\bibitem{Bhagat_2025_arxiv_conditionalconsistencyguidedimage}
\bibinfo{author}{Bhagat, A.}, \bibinfo{author}{Jain, M.} \& \bibinfo{author}{Subramanyam, A.~V.}
\newblock \bibinfo{title}{Conditional consistency guided image translation and enhancement} (\bibinfo{year}{2025}).
\newblock \urlprefix\url{https://arxiv.org/abs/2501.01223}.
\newblock \eprint{2501.01223}.

\bibitem{Li_2023_CVPR_BBDM}
\bibinfo{author}{Li, B.}, \bibinfo{author}{Xue, K.}, \bibinfo{author}{Liu, B.} \& \bibinfo{author}{Lai, Y.-K.}
\newblock \bibinfo{title}{Bbdm: Image-to-image translation with brownian bridge diffusion models}.
\newblock In \emph{\bibinfo{booktitle}{Proceedings of the IEEE/CVF Conference on Computer Vision and Pattern Recognition (CVPR)}}, \bibinfo{pages}{1952--1961} (\bibinfo{year}{2023}).

\bibitem{Zhang_2025_NatureComms_BBDM}
\bibinfo{author}{Zhang, Y.} \emph{et~al.}
\newblock \bibinfo{title}{Pixel super-resolved virtual staining of label-free tissue using diffusion models}.
\newblock \emph{\bibinfo{journal}{Nature Communications}} \textbf{\bibinfo{volume}{16}} (\bibinfo{year}{2025}).
\newblock \urlprefix\url{http://dx.doi.org/10.1038/s41467-025-60387-z}.

\bibitem{Wodzinski_2024_arxiv_DeeperHistReg}
\bibinfo{author}{Wodzinski, M.}, \bibinfo{author}{Marini, N.}, \bibinfo{author}{Atzori, M.} \& \bibinfo{author}{Müller, H.}
\newblock \bibinfo{title}{Deeperhistreg: Robust whole slide images registration framework} (\bibinfo{year}{2024}).
\newblock \urlprefix\url{https://arxiv.org/abs/2404.14434}.
\newblock \eprint{2404.14434}.

\bibitem{Chiaruttini_2022_FrontiersCS_Warpy}
\bibinfo{author}{Chiaruttini, N.} \emph{et~al.}
\newblock \bibinfo{title}{An open-source whole slide image registration workflow at cellular precision using fiji, qupath and elastix}.
\newblock \emph{\bibinfo{journal}{Frontiers in Computer Science}} \textbf{\bibinfo{volume}{3}} (\bibinfo{year}{2022}).
\newblock \urlprefix\url{http://dx.doi.org/10.3389/fcomp.2021.780026}.

\bibitem{Bangare_2015_IJAER_otsu}
\bibinfo{author}{Bangare, S.~L.}, \bibinfo{author}{Dubal, A.}, \bibinfo{author}{Bangare, P.~S.} \& \bibinfo{author}{Patil, S.}
\newblock \bibinfo{title}{Reviewing otsu’s method for image thresholding}.
\newblock \emph{\bibinfo{journal}{International Journal of Applied Engineering Research}} \textbf{\bibinfo{volume}{10}}, \bibinfo{pages}{21777–21783} (\bibinfo{year}{2015}).
\newblock \urlprefix\url{http://dx.doi.org/10.37622/IJAER/10.9.2015.21777-21783}.

\bibitem{pp2p_git}
\bibinfo{author}{{BUPT AI CZ Group}}.
\newblock \bibinfo{title}{Bci}.
\newblock \bibinfo{howpublished}{\url{https://github.com/bupt-ai-cz/BCI}} (\bibinfo{year}{2023}).

\bibitem{asp_git}
\bibinfo{author}{Li, F.}
\newblock \bibinfo{title}{Adaptive supervised patchnce}.
\newblock \bibinfo{howpublished}{\url{https://github.com/lifangda01/ AdaptiveSupervisedPatchNCE}} (\bibinfo{year}{2021}).

\bibitem{bcistainer_git}
\bibinfo{author}{Qu, Q.}
\newblock \bibinfo{title}{Bcistainer}.
\newblock \bibinfo{howpublished}{\url{https://github.com/quqixun/BCIStainer}} (\bibinfo{year}{2021}).

\bibitem{ddib_git}
\bibinfo{author}{Su, X.}
\newblock \bibinfo{title}{Ddib: Dual-domain information bottleneck for stain transfer}.
\newblock \bibinfo{howpublished}{\url{https://github.com/suxuann/ddib}} (\bibinfo{year}{2023}).

\bibitem{cm_git}
\bibinfo{author}{Bhagat, A.}
\newblock \bibinfo{title}{Conditional consistency models}.
\newblock \bibinfo{howpublished}{\url{https://github.com/amilbhagat/Conditional-Consistency-Models}} (\bibinfo{year}{2023}).

\bibitem{Hore_2010_CPR_PSNR}
\bibinfo{author}{Horé, A.} \& \bibinfo{author}{Ziou, D.}
\newblock \bibinfo{title}{Image quality metrics: Psnr vs. ssim}.
\newblock In \emph{\bibinfo{booktitle}{2010 20th International Conference on Pattern Recognition}}, \bibinfo{pages}{2366--2369} (\bibinfo{year}{2010}).

\bibitem{Heusel_2017_arxiv_FID}
\bibinfo{author}{Heusel, M.} \emph{et~al.}
\newblock \bibinfo{title}{Gans trained by a two time-scale update rule converge to a nash equilibrium}.
\newblock \emph{\bibinfo{journal}{arXiv}}  (\bibinfo{year}{2017}).

\bibitem{Bińkowski_2021_arxiv_KID)}
\bibinfo{author}{Bińkowski, M.}, \bibinfo{author}{Sutherland, D.~J.}, \bibinfo{author}{Arbel, M.} \& \bibinfo{author}{Gretton, A.}
\newblock \bibinfo{title}{Demystifying mmd gans} (\bibinfo{year}{2021}).
\newblock \urlprefix\url{https://arxiv.org/abs/1801.01401}.
\newblock \eprint{1801.01401}.

\bibitem{Zhang_2018_CVPR_LPIPS}
\bibinfo{author}{Zhang, R.}, \bibinfo{author}{Isola, P.}, \bibinfo{author}{Efros, A.~A.}, \bibinfo{author}{Shechtman, E.} \& \bibinfo{author}{Wang, O.}
\newblock \bibinfo{title}{The unreasonable effectiveness of deep features as a perceptual metric}.
\newblock In \emph{\bibinfo{booktitle}{2018 IEEE/CVF Conference on Computer Vision and Pattern Recognition}}, \bibinfo{pages}{586--595} (\bibinfo{year}{2018}).

\bibitem{Bates_2015_JSS_lme4}
\bibinfo{author}{Bates, D.}, \bibinfo{author}{Mächler, M.}, \bibinfo{author}{Bolker, B.} \& \bibinfo{author}{Walker, S.}
\newblock \bibinfo{title}{Fitting linear mixed-effects models using lme4}.
\newblock \emph{\bibinfo{journal}{Journal of Statistical Software}} \textbf{\bibinfo{volume}{67}}, \bibinfo{pages}{1–48} (\bibinfo{year}{2015}).
\newblock \urlprefix\url{https://www.jstatsoft.org/index.php/jss/article/view/v067i01}.

\end{thebibliography}

\end{document}